# DESIGN STUDIO 2.0: AUGMENTING REFLECTIVE ARCHITECTURAL DESIGN LEARNING

*Dr. Burak Pak,*
*Sint-Lucas School of Architecture, Faculty of Architecture and Arts, Hogeschool voor W&K, Association KU Leuven;*
*burak.pak@architectuur.sintlucas.wenk.be  http://blog.associatie.kuleuven.be/burakpak/*

*Prof. Dr. Johan Verbeke,*
*Sint-Lucas School of Architecture, Faculty of Architecture and Arts, Hogeschool voor W&K, Association KU Leuven;*
*johan.verbeke@architectuur.sintlucas.wenk.be*

**SUMMARY:** Web 2.0 is beyond a jargon describing technological transformation: it refers to new strategies, tools and techniques that encourage and augment informed, creative and social inter(actions). When considered in an educational context, Web 2.0 provides various opportunities for enhanced integration and for improving the learning processes in information-rich collaborative disciplines such as urban planning and architectural design. The dialogue between the design students and studio teachers can be mediated in various ways by creating novel learning spaces using Web 2.0-based social software and information aggregation services; and brought to a level where the Web 2.0 environment supports, augments and enriches the reflective learning processes. We propose to call this new setting **"Design Studio 2.0"**. We suggest that Design Studio 2.0 can provide numerous opportunities which are not fully or easily available in a conventional design studio setting. In this context, we will introduce a web-based geographic virtual environment model (GEO-VEM) and discuss how we reconfigured and rescaled this model with the objective of supporting an international urban design studio by encouraging students to make a collaborative and location-based analysis of a project site (the Brussels-Charleroi Canal). Pursuing the discussion further, we will present our experiences and observations of this design studio including web use statistics, and the results of student attitude surveys. In conclusion, we will reflect the difficulties and challenges of using the GEO-VEM in the Design Studio in a blended learning context and develop future prospects. As a result, we will introduce a set of key criteria for the development and implementation of an effective e-learning environment as a sustainable platform for supporting the Design Studio 2.0.

**KEYWORDS:** Architectural design education; e-learning; blended learning; social web application; Design Studio

# 1. INTRODUCTION

The last decade has witnessed the proliferation of new web-based social software and information aggregation services which facilitate social knowledge construction. These are commonly put under the umbrella of the term "Web 2.0" which has been described in the manifesto of O'Reilly issued in 2005 as *"practices in which web is used as a platform for harnessing collective intelligence, delivered as a service not a product, based on lightweight programming models, backed by a specialized database, supporting PC and non-PC devices and providing a rich user experience"* (http://oreilly.com/pub/a/web2/archive/what-is-web-20.html).

Web 2.0 practices can be considered novel in ways that they enable *learning as a social process* (Brown and Adler, 2008) by creation of rich content through discussion, reflection and informed consensus. In this context, Web 2.0 is beyond a jargon or a "buzzword" describing technological transformation: it refers to new strategies, tools and techniques that encourage and augment informed, creative and social inter(actions).

In this paper, we claim that these new strategies, tools and techniques can be used as a basis for constructivist learning; especially in highly reflective fields such as architectural and urban design education. Web 2.0-based social software and information aggregation services can be utilized as additional platforms to mediate and extend the reflective conversation between the teachers and students in the architectural design studios. This inevitably involves combining online and face-to-face activities, in other words, enabling blended-learning (Heinze and Procter, 2004). We name this new setting **"Design Studio 2.0"** and suggest that it can improve the design learning experience by helping the students to develop a deeper understanding of the materials at hand, motivate them to learn from other students' works and improve the quality of their designs.

Following these motivations, we will begin our paper by reviewing recent valuable applications of e-learning 2.0 strategies in the design studio (Section 2). We will discuss the conventional position of the design studio in the curriculum of Schools of Architecture (and especially in Sint-Lucas School of Architecture) and how this curriculum was redesigned to support dynamic and sustainable education based on an interdisciplinary learning process. A crucial element here will be the interaction between the design studio and more theoretical subjects. In order to extend this discussion, we will include the experiences from the OIKODOMOS project in which a Web 2.0 platform was created to support the learning of students from different schools of architecture and urban design in Europe.

In Section 3, we will introduce the web-based geographic virtual environment model (GEO-VEM) that we developed in the framework of a long-term research project. In brief, this environment is a Web 2.0 application hybrid specifically developed for the representation and communication of alternative urban development projects for the Brussels-Capital Region. It combines Semantic MediaWiki and Google Earth API for representing textual data, imagery, concepts maps, 3D models and time-based information in a geolocated format. Following the introduction of the GEO-VEM, we will express how we reconfigured and rescaled this model with the aim of encouraging students to make a collaborative and location-based analysis of the project area (the Brussels-Charleroi Canal) and share their findings with each other.

In Section 4, we will present our experiences from an e-learning enabled urban design studio using illustrations and sample student works. This discussion will be elaborated with web use statistics and the results of the student attitude survey including the student satisfaction questionnaire (Section 5).

In conclusion (Section 6), we will reflect the difficulties and challenges of using the GEO-VEM in the Design Studio and develop future prospects.

# 2. REDESIGNING THE DESIGN STUDIO: AN E-LEARNING ENABLED STUDIO

The structure of curricula in Schools of Architecture can be traced back to the Vitruvian triad "firmitas, utilitas, venustas" which implies that different perspectives and a variety of competences are essential for graduates in the field of architecture. Until the mid-nineteenth century, architectural education was based on an apprentice system where young architects served under the mastery of an accomplished architect (Jacques, 1982).

The early roots of the architectural design studio are often referenced to the nineteenth century *ateliers* of the *École des Beaux-Arts* located in Paris. In *École des Beaux-Arts*, the students were offered courses and *ateliers* on various subjects. The *ateliers* brought a new approach to the architectural design education which can be described as "learning by doing" (Schön, 1987). Since then, the design studio has been the core of the education in the field of Architecture (as it is even more in arts and design). The design knowledge, thinking and understanding created in the design studio and the experience and knowledge that is transferred from practice has been essential to the field. This knowledge is mostly transferred in a tacit way through projects, *charettes*, discussions, workshops and other activities.

In the 1920s, with the influence of the modernist movement, architectural education was reformed to fit the needs of the emerging socio-economical context. The heart of the modernist movement, the German Bauhaus School led this transformation and integration of new concepts related to mass production and new technologies. This reform has made a significant and global impact on the schools of architecture, especially during and after the Second World War. Although the Bauhaus ideas have transformed the architectural education, the studio-based learning model remained mostly unchanged (Lackey, 1999).

If we look at the situation today, we see that the design studio (where the designing competence is educated) still takes a central role in architectural education. According to the European Association for Architectural Education (EAAE), the curricula of schools of architecture include between 25% (in the more engineering oriented schools) to 60% of design studio activities (http://www.eaae.be/members_new.php). This applies to undergraduate as well as graduate education. Moreover, most schools have a large group of staff who combine their academic activities with architectural practice. This is very similar to the arts, where most of the staff are active artists. The input from these practices is crucial for the development of the field and architectural education usually still is situated in a strong master-apprentice relation.

Accreditation panels regularly report a split between design studio knowledge and knowledge transferred in other courses and see this as a problem. In order to respond to this challenge, the Sint-Lucas School of Architecture decided in 2003 to develop a strong interaction between the design studio teaching and the courses with a more theoretical stance. It developed the concept of design tracks where each design studio is complemented by two theoretical components. Design tasks as well as the content and focus of both theoretical components are jointly prepared by the staff involved (usually one design studio teaching staff member and two teaching staff members with a different background). Although the initial phase to implement the design tracks was difficult, soon it turned out that staff as well as students reported very positive experiences.

## 2.1 E-learning, Blended Learning and Design Studios

Design studio involves the collective construction of knowledge through rigorous dialogue between designers/teachers and the students and teaching/learning through reflection-in-action (Schön, 1983). In this sense, constructivist epistemology provides a fruitful framework to describe the existing and develop new knowledge created and recreated through dialogue and shared experiences during the design studio. All of these experiences accumulate in what we know when making a new design and this new design helps us to deepen and extend our understanding, hence contributing to our body of knowledge and mental space (Young, 1994).

In close relation with the constructivist theoretical framework, an e-learning enabled design studio (in contrast with the conventional design studio), promotes "community building" and "social learning" rather than one-on-one and face-to-face communication. With the help of e-learning environments, students' time spent outside the classes can be more efficiently and effectively valorized (Shirky, 2010). While they are disconnected from the physical studio environment, the students can still learn from and comment on each others' projects and create a collective understanding of the design problem(s), the design context and the whole studio process. In addition, the course materials and various design products that are created during the design studio can be documented in a structured manner and transferred to concurrent and future design studios, designers and design researchers in various geographies. It is important to keep in mind that meaning is not implicit in the structured information; rather, learners (both students and teachers) should assign meaning to it (Puntambekar and Young, 2003). Therefore, the focus of an e-learning enabled design studio should not solely be on the documentation and structuring of design information but on the collective construction of understanding and knowledge.

Besides the theories on constructivist learning, various practical studies have been made in the past to test the potentials of using e-learning platforms in the design studio. For instance, Lindquist (2006) made a case study of the use of Wikis by students in a landscape architecture design studio. He noted that being one of the most powerful collaborative authoring tools, Wikis are strong social Web 2.0 environments that can be used as a companion to design studios. As a result of his study, he found that Wikis are effective means of teacher and student collaboration, especially in the early stages of a landscape design project. In line with these findings, Burrow and Burry (2006) reported the effective use of Wikis as an internationally distributed design research network incorporating diverse forms of expertise. They have used a Wiki for collaborative design; as a facilitating platform to reflectively connect the architect on site (at the Sagrada Família Church in Barcelona) and other members of design team who continue to work on the project in Australia. Similarly, Chase et al. (2008) introduced the "Wikitecture" concept as a decentralized method of open source co-production and tested the use of a 3D Wiki to collaboratively develop a design competition entry. They observed that this Web 2.0 environment allowed *"increased opportunity to self-select and self-organize around projects that interest the designers most, with increased benefits of creativity, motivation, and flexibility, resulting in an altogether more efficient design process"*.

Considering all of the findings reported above, it would not be wrong to state that e-learning does not entirely replace face-to-face communication in the design studio; rather it augments the whole learning process. The success of the e-learning practices lies in how well they mix and balance face-to-face and e-learning activities; or in other words how they support *blended learning*. With the help of Web 2.0 technologies, blended learning enables the institutions to stick with their existing academic curricula and supports the design and implementation of learning activities in collaboration (Madrazo et al, 2010).

## 2.2    Experiences from the OIKODOMOS Research Project

In view of the above perspectives, the combination of constructivist learning in architectural education with the possibilities of Information and Communication Technologies (ICT) and especially Web 2.0 results in fascinating learning processes. Through online activities and workshop activities as well as face-to-face activities (a blended learning environment), students and teaching staff can construct their understanding of the design problem, explore possibilities, develop solutions and communicate their insight and knowledge. A good example of this type of learning using Web 2.0 technologies is the OIKODOMOS project.

As described by Madrazo and Riddy (2011), the OIKODOMOS Virtual Campus -an EC funded project to which one of the authors has contributed- is a learning space where teachers and students of schools of architecture and urban planning collaborate in the design and implementation of learning activities dedicated to the study of housing in contemporary Europe. In this space, the learning experience of the students is shaped through an intertwining of on-line and on-site activities or "a blended learning" approach. Moreover, the technological platform is specifically designed to stimulate sharing and developing of experience, understanding, insight and knowledge by and from the participants. The platform consists of a series of learning tasks and attempts to stimulate learners to interact, share, discuss and develop their understanding. Hence, the OIKODOMOS virtual campus is a good example of Web 2.0 learning as social interaction and discussions are core processes of the learning. The activities on the virtual campus are complemented by real-life teaching and workshops to help the integration of the understanding and knowledge of the students (Verbeke et al., 2012).

The result of the learning is not only the designs made by students in the design studios, but also the increased understanding developed during undertaking these learning activities. The understanding and insight of the learning also contributes to the overall body of knowledge which is extended and constructed from the experience.

## 2.3    The Design Studio 2.0 concept

The discussion above clearly illustrates that Web 2.0 tools and strategies can enhance learning in the design studios by enriching the reflective learning processes. As referenced in the introduction, we call this new setting **"Design Studio 2.0"**. Design Studio 2.0 differs from the conventional design studio in terms of *available communication modes and styles, learning experiences, studio focus, studio environment, time, information resources and representation of design information* (Table 1). It offers numerous opportunities which are not fully or easily available in a conventional design studio setting.

First of all, it can promote and facilitate *reflective learning-in-action* in a novel pedagogical context, in which various communication modes and styles are supported. The possibility of one-to-one, one-to-many, many-to-one and many-to-many communication allows more flexible and adaptable interactions and number of design students. Furthermore, in the new setting (as illustrated by the OIKODOMOS project and our own experiences as described in section 4), design studio learning is complemented by asynchronous activities in the virtual campus. In this sense, the new e-learning environments are extending the design studio learning in place and time. They offer the learners the possibility of extended online discussions complementing the activities in the design studio. In contrast, the discussions in the conventional design studio take place in small groups, complemented by plenaries and reviews.

Face-to-face activities of the conventional design studio are brought to another level by integrating Web 2.0 technologies in the learning environment. As will be shown later, these environments improve the overall learning experience and help the student to develop a deeper understanding of the materials at hand.

*TABLE 1: Comparison of Conventional Design Studio and Design Studio 2.0.*

|  | **Conventional Design Studio** | **Design Studio 2.0** |
|---|---|---|
| **Communication Modes** | One-to-one / One-to-many | One-to-one / One-to-many / Many to one / Many-to-many |
| **Communication Forms** | Synchronous | Synchronous, Asynchronous, Combined |
| **Communication Styles** | Face-to-face | Face-to-face, Avatar-to-Avatar |
| **Learning Type** | Conventional design studio learning | Design studio complemented with blended learning |
| **Focus** | Studio coordinator, design products | Students, design products and the learning processes |
| **Studio Environment** | Physical (Local) | Physical and Virtual (Global) |
| **Time** | Studio hours and individual studies | Studio hours and individual studies: asynchronous learning (mediated) |
| **Representation of Design Information** | Sketches, drawings, maps and models, printed versions of computer drawings, renderings etc. | 3D models (4D with the inclusion of time), scanned versions of sketches and drawings (2D), digital versions of computer drawings and renderings, dynamic maps, geolocated notes, comments etc. |

## 3. THE WEB-BASED GEOGRAPHIC VIRTUAL ENVIRONMENT MODEL (GEO-VEM)

Web-based Geographic Virtual Environments can be briefly defined as Web 2.0 applications that combine various types of data, geographic information services and functionalities from different sources. Examples of these are GMapCreator and Maptube, both developed at University College London CASA Centre (Hudson-Smith et al., 2009).

In this section, we will present a Web-based Geographic Virtual Environment Model (GEO-VEM) that we developed in the framework of a long-term research project supported by the Brussels Innovation and Research Institute. Accordingly, we will explain how we redesigned this environment with the aim of encouraging students to make a collaborative, open source and location-based analysis of the project area and share their findings with each other.

### 3.1 The Original Virtual Environment Model Developed for the Brussels Capital Region: Revealing the Context, Aims, Motivations and Design

Numerous urban designers, planners and institutions have created various urban development projects to improve the city and the residents' quality of life in the Brussels Capital Region. A plenty of those projects were partially or fully realized, while a significant number of projects are still waiting to be implemented or were only intended to provide conceptual ideas. In this paper, we will call the latter **"alternative urban development** **projects".** An alternative urban development project possesses a performative power as a contribution to redefine reality and a social activity (Pak and Kuhk, 2009). This power can only be unlocked with the tools that enable the planning actors to retrieve, discuss and share projects efficiently and effectively; and therefore contribute to their development in a constructive manner.

In the context of Brussels, UrbIS Geographical Information System is the main tool used by the authorities for storage and distribution of geographic data. This system majorly focuses on keeping an inventory of the existing "**reality**" and physical condition of the city on different scales. In contrast, *alternative urban development projects* are representations of the "**imaginary**" and they focus on different possible futures instead of single and accurate representations.

With the motivations above, we have initiated a research project at the beginning of 2009 for creating a GEO-VEM for the communication, analysis and deliberation of alternative urban development projects specifically designed for the Brussels Capital Region. Our aim was to design and develop a digital environment through which planning actors can learn, exchange ideas and shape the future strategies. In accordance, we have conducted an in-depth analysis of the alternative urban development projects prepared for the Brussels Capital Region to determine their characteristics and collect the types of information they contain. The results of the analysis showed that alternative urban development projects include the following types of information:

- Objectives, hypotheses, decisions related to strategies at different levels, planning, tactics, limits, design, implementation; descriptions, definitions and methods relating to a variety of concepts and themes *(Textual information)*
- Schedules and timelines - provisional planning *(Temporal information)*
- Photos, sketches, mock-ups, schemes, maps, plans, section and perspective drawings *(Two dimensional design data in raster and vector format)*
- Physical and virtual 3D models *(Three dimensional data)*

As a result, based on the studies above and consultations with the *Agency for Territorial Development (ADT/ATO), Brussels Environment Organization (BRAL)* and *Brussels Informatics Center (CIRB)* a general framework was designed as a Web 2.0 application hybrid based on a combination of different representations and carefully organized into two parts (Figure 1). The interface on the left offers a *Google Earth API* visualization window and an integrated *time-based map*, specifically addressing the geographical location. A geographic wiki serves as a basis for integrating and linking to different types of content which will be interpreted and rendered by various applications. The interface on the right foresees *an interactive concept map* and a wiki-based hypertext window that serves textual data and images with search functionality.

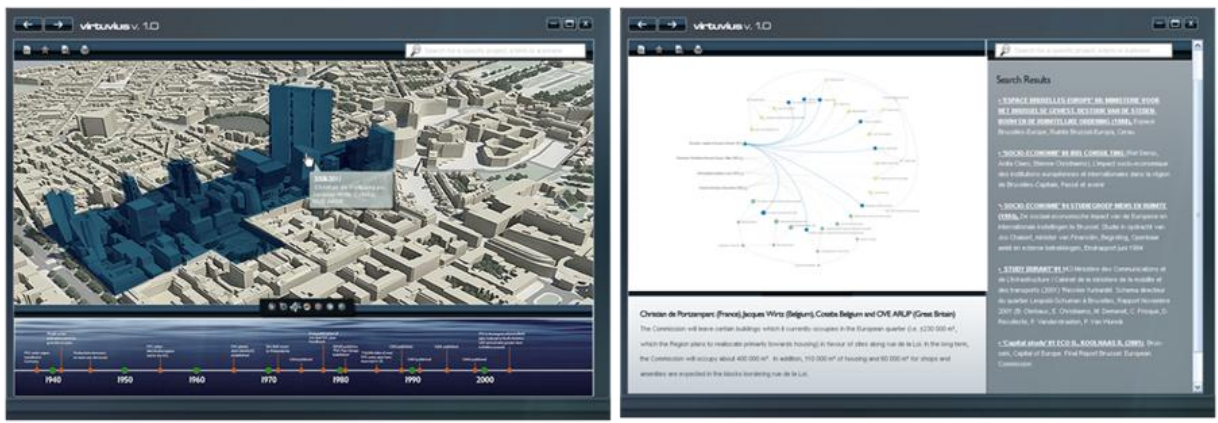

*FIG. 1: Outline of the Web 2.0 based GEO-VEM for communication, analysis and deliberation of alternative urban development projects (AUDPs).*

These functionalities refer to certain types of information related to different attributes of the alternative urban development projects. Google Earth API is used to visualize *georeferenced 3D models* and *textual data* synchronized and controlled by the *timeline* located at the bottom (Figure 1). Google API provides relatively *high resolution aerial imagery* and *3D city models* at an unrivalled extent and allows adding and visualizing user generated content through a highly customizable interface. KML (Keyhole Markup Language) schema, used by this API, facilitates attributes that allow effective visualization of geographic data.

In this virtual environment model, Google Maps API is also embedded in this system via "Google Maps MediaWiki Extension" (http://www.mediawiki.org/wiki/Extension:Google_Maps) developed by Evan Miller in 2009. The semantic mapping functionality is made available through "Semantic Maps Extension" developed by de Dauw et al. in 2009 (http://www.mediawiki.org/wiki/Extension:Semantic_Maps) and the timelines and concept maps are connected to related "SIMILE" and "FLARE" visualization libraries by Semantic Results Formats Extension developed by Dengler et al. in 2010 (http://www.mediawiki.org/wiki/Extension:Semantic_Result_Formats)(Figure 2).

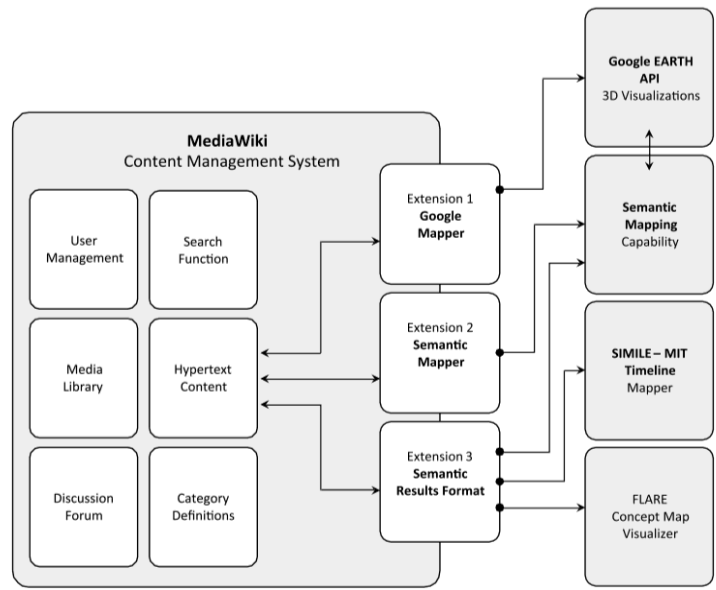

*FIG. 2: The structure of the GEO-VEM developed for the Brussels Capital Region.*

## 3.2 The Rescaled Version of the GEO-VEM Developed for an International Graduate Urban Design Studio organized in Brussels

In order to evaluate the Design Studio 2.0 concept in real life, we have specifically rescaled and customized the Web 2.0 based geographic virtual environment that we have developed for the Brussels Capital Region. We actively got involved into a graduate design studio at the Sint-Lucas School of Architecture during the first eight weeks of the fall semester of 2010 and worked in close collaboration with the studio coordinators.

We created a custom setup to encourage the students to make a collaborative, open source and location-based analysis of the project site and share their findings with each other. Therefore, we divided them in groups of three to five students and asked each team to define a specific theme (these themes were later used by the students to organize their works). Because of the open structure of the MediaWiki, students were enabled to collaboratively edit each other's contributions.

In addition, the main interface has been reorganized to support and fit in the design studio setting (Figure 3). On the home page, besides announcements, events and a brief description of the design studio, we have placed a collaborative interactive Google Map that shows the intervention zones that are defined by the students.

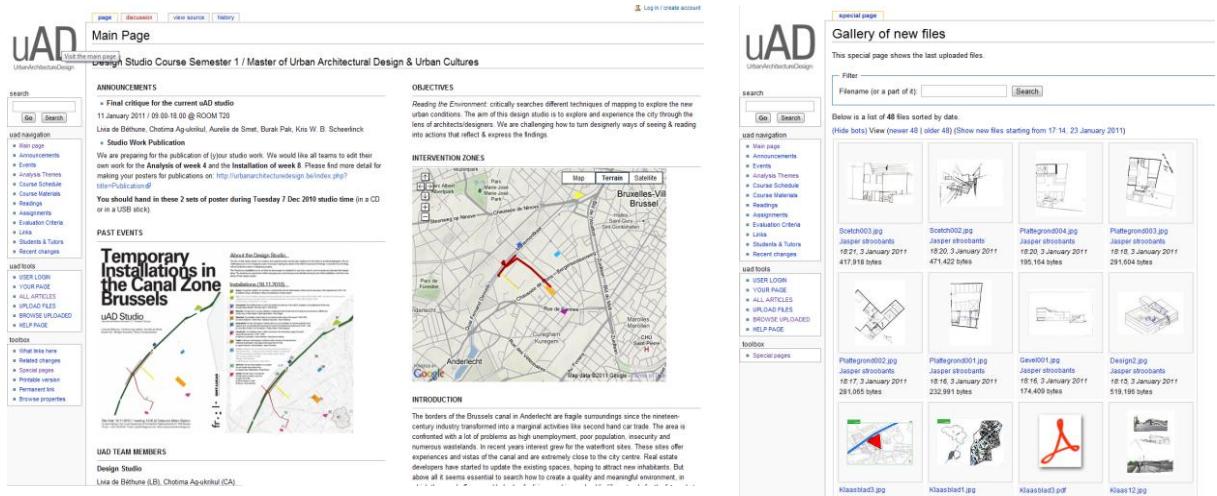

*FIG. 3: The interface of the rescaled version of the Web 2.0-based GEO-VEM used in the design studio: the main page and the gallery of uploaded files.*

The whole content framework was organized around the eleven themes that were assigned to each student group: activity, density, consumption, networks, (in)formal, interaction, borders-limits, everyday life, fragile, contrast and time. Student groups were asked to create a new page every week related to the relevant theme. By this way, we were able to monitor the development of the analysis process and new design ideas,

Moreover, a Google Maps extension was enabled to allow the students to geolocate a certain area, mark and initiate a new topic on a specific place, draw and comment on the dynamic map (Figure 4). We have provided the students with the opportunity to import and export KML files and superpose multiple maps to create a new one. It was also possible to place more than one map on a single page, a feature that can be utilized for the comparative evaluation alternatives.

The rescaled version of the GEO-VEM also included the Semantic MediaWiki extension that enables the codification of the semantic relations through the regular editing interface.

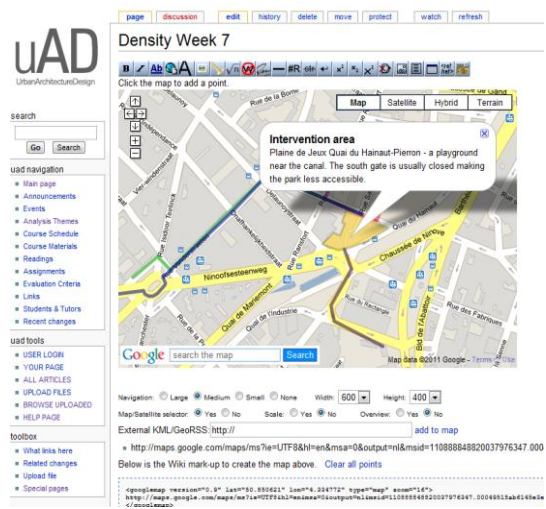

*FIG. 4: Google Maps extension was enabled to allow the students to geolocate a certain area, mark and initiate a new topic on a specific place, draw and comment*

In order to facilitate the use of these complex functionalities, we have prepared special online help tutorials on how to upload and edit text, links and images using wiki interface and create and edit maps by the help of Google maps extension. The classical wiki "cheat sheet" was also made available on our help page.

Furthermore, hands-on workshops were held every week to introduce the Web 2.0-based GEO-VEM, promote its use and receive direct feedback from the students. These workshops included basic concepts on web based collaborative virtual environments, interface use and practice sessions.

## 4. EXPERIENCES FROM AN E-LEARNING ENABLED GRADUATE URBAN DESIGN STUDIO (2.0)

As introduced in the previous section, we tested the rescaled version of the GEO-VEM in a graduate urban architectural design studio (uAD).

The design studio was especially configured to test the use of this rescaled web-based geographic virtual environment for the analysis of the urban setting. This studio included 39 students of which 36 were ERASMUS exchange students coming from all over Europe and Japan. A majority of these students had no prior knowledge of Brussels. Therefore, it was necessary for the students to develop a thorough understanding of the social and spatial characteristics of the city.

The GEO-VEM was actively used between the first and eighth weeks of the design studio with a focus on the analysis of the project site and developing a preliminary design. The students worked in groups throughout this phase, sharing information and their findings with each other. After the eighth week, the students used their experiences to create a temporary installation on the project area and established a reflective communication with the inhabitants. After this phase, they developed an urban design project on the same area considering all of their experiences.

The studio also included field tours, tutor lectures, three workshops and interactions with the theoretical component course that promotes the integration of theoretical concepts into the design process. The web-based geographic virtual environment was also actively used during these activities (Figure 5, on the left). For instance, during the field tours, the whole travel route was recorded by a GPS device and then transformed into KML format and joined with the student photos in a geolocated format on the GEO-VEM.

Similarly, during the neogeography workshop, the students made individual hand sketches reflecting their experiences of the city and layered them over a dynamic Google Map to create a map of collective knowledge (Figure 5, on the right).

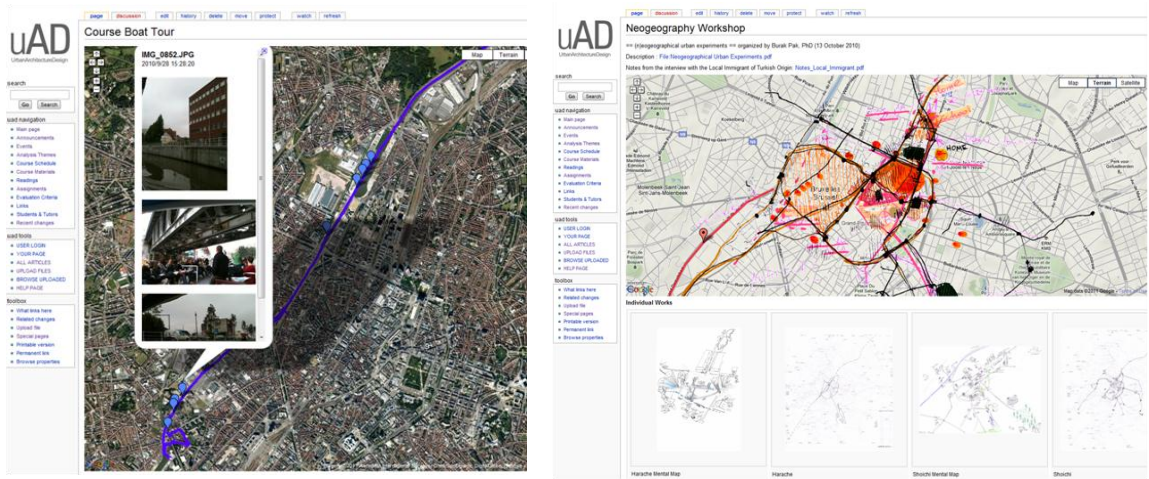

*FIG. 5: Various opportunities provided by the Web 2.0 environments were used during the field trips: GPS route of the field trip and geolocated student photos (on the left), layering the scanned hand sketches of the students over a dynamic Google Map during the neogeography workshop to create a map of collective knowledge (on the right).*

Considering the modality of the student activities, the design studio setup can be called a blended-learning environment. The weekly critique of the student works took place in a regular studio, in a conventional face-to-face manner. A video projector was used for most of the group presentations. In various instances, the students used the web-based geographic virtual environment to refer and present a past finding or a design idea.

Among the online activities were regularly uploading their findings and ideas on their individual pilot project to their group theme pages (in a collaborative manner); geolocating and describing a small project that is meaningful to their personal experiences (on their individual wiki user pages) and reading and discussion of the tutor week seminars.

During the whole studio process the students also actively used the GEO-VEM in a reflective manner, and created an impressive online inventory with 66 topics (pages), organized according to 11 themes. These topics included various analysis findings, sketches, photos, maps, studio presentations and texts describing their experiences and thoughts on their future projects (Figure 6). The total size of the uploaded data was around five gigabytes. The contents of the group pages were not moderated by the tutors, but they had to include: a verbal description of group findings (linked with maps), photos of (physical) models, sketches and the PDF version of group presentations and/or posters (Figure 6).

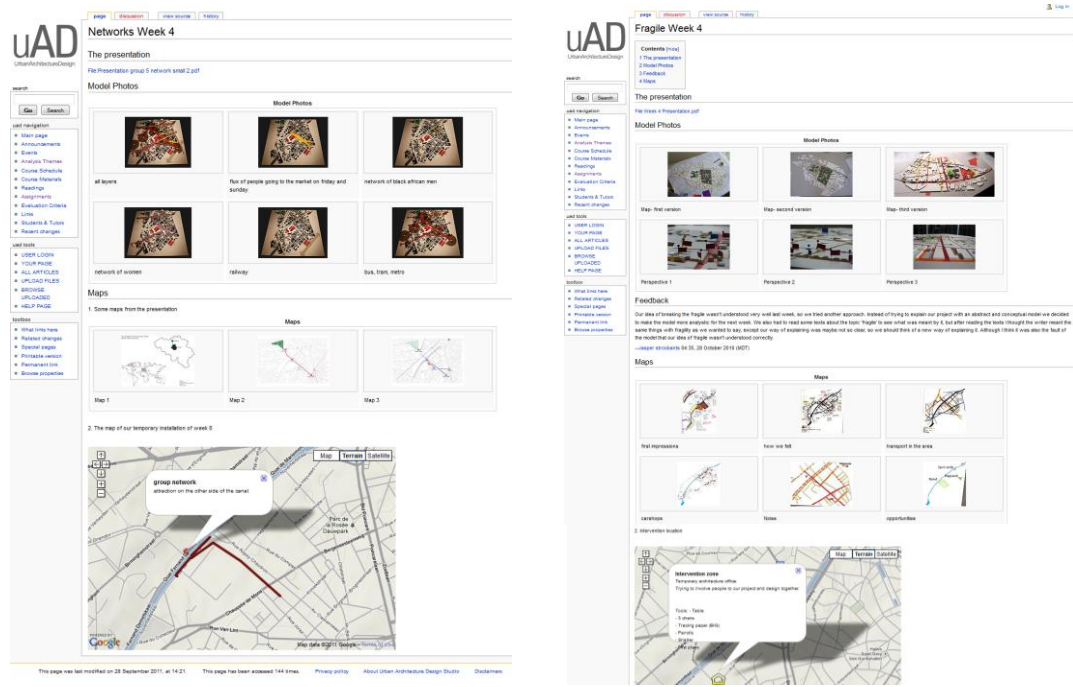

*FIG. 6: Screenshots from the collaborative analysis of the project site and design ideas produced by the students. The students have uploaded their model photos, diagrams and geolocated their findings (The view on the left has been cropped to fit in the format).*

The groups uploaded their studies weekly until the end of the eighth week. This allowed the regular monitoring of the student works, and most important, tracking their weekly progress which was used as the main indicator for evaluating their success. The Web 2.0-based GEO-VEM motivated the students to construct collaborative design diaries in a structured format which has been found to be easy to evaluate by the studio coordinators and invited lecturers. Moreover, these contents were also open to other studio members and organizers including the international audience, though the level of interaction is difficult to measure except for the number of page views.

## 5. EVALUATION OF THE URBAN DESIGN STUDIO 2.0 EXPERIMENT

We have employed a variety of methods to evaluate our Design Studio 2.0 experiment and gather information on the nature and intensity of the students' online collaboration. Among these were the on-site web analytics, a student attitude survey, a post task user satisfaction questionnaire and feedback meetings. In this section, we are going to discuss these observations in combination with our individual experiences during the design studio and student performances (grades).

## 5.1 Web Analytics

According to the web analytics results, the 39 users created a total of 763 pages (including talk pages, redirects, etc) and uploaded 617 files to the system. The total number of page edits was 2444 and average number of edits per page was 3.20. If we exclusively look at the total number of group project theme pages, we see that the students contributed to 66 pages. 48 out of 66 pages (79%) were edited by more than one student (Figure 7). We can conclude from this observation that the students created content together through the virtual environment during the design studio. This observation is also in line with our informal perception that working in groups motivated them to share information and findings and construct a collective memory of their project sites.

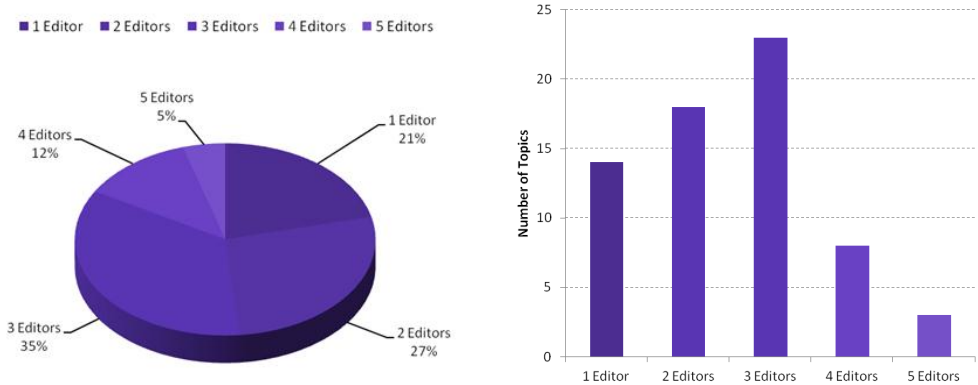

FIG. 7: Percentage of pages according to the number of students that contributed to them. 79 % of the topics (pages) were edited by more than one student.

When we analyzed the contributions on an individual basis, we found that the students made an average of 38.89 edits during the whole studio (Figure 8). The majority (26) of the students have made 1 to 50 edits whereas only 3 students made more than 100 edits. These numbers and figure 8 show the distribution of students' cumulative contribution.

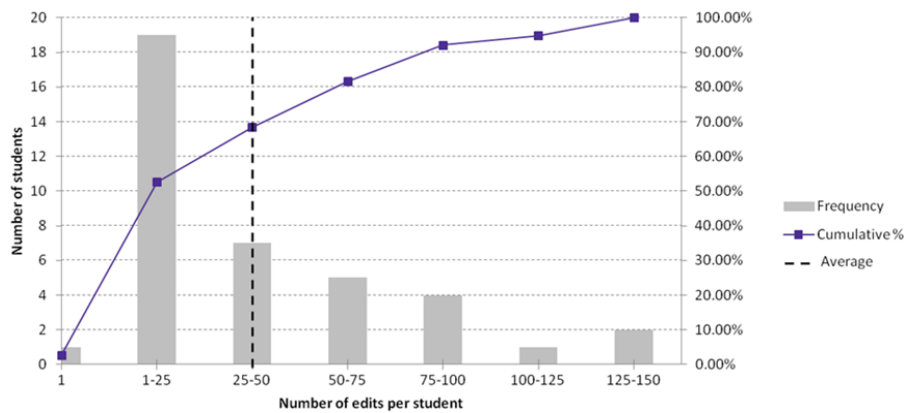

FIG. 8: Histogram of individual student edits. Average edits per student is 38.89.

Considering the distribution of individual edits, an important subject to address was the possible differences in gender. Glott et al. (2010) have made a study on the differences in the gender composition of readers and contributors of Wikipedia and found that only 12.64 percent of contributors were female (this result may also be caused by the false representation of user profiles). In accordance with this study, we compared the average number of edits by female and male students during our design studio experiment. Although average contribution per female students were slightly lower (38.19) than the male students (40.5) we did not detect any significant gender difference.

As a final study for the evaluation of the web analytics and the design studio experiment, we compared the student grades at the end of the analysis period with the total number of group edits (Figure 9). The idea behind this comparative analysis was that there should possibly be some relation between the intensive collaborative use of the environment and group success (grades) in the analysis phase of our design studio (in week eight they were evaluated as a group, not individually).

As a result, we found that the groups who received higher grades had made relatively more collaborative edits (and vice versa) (Figure 9). It is impossible to derive a direct causality out of this finding, but the data points indicate a possible correlation between collaborative edits and student performance. When combined with our design studio observations, these findings suggest that collaborative use of the proposed virtual environment as a knowledge resource may improve the performance of the student groups (i.e. "Everyday Life" group). On the other hand, after a threshold (around 240 edits in this case), focusing too much on the web environment may also decrease the performance of the students (i.e. "Fragile" and "Networks" groups). Of course, we must note that these observations are highly dependent on the profile and individual characteristics of the students.

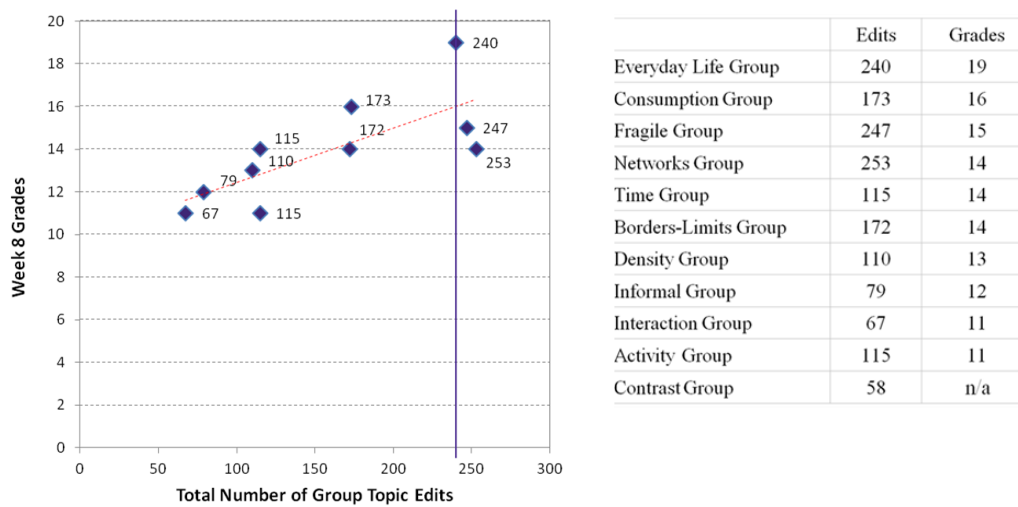

FIG. 9: Scatter graph of Student Groups' grades compared with the total number of student edits as a group (exclusively the group project theme pages).

## 5.2 The Student Attitude Survey

We have conducted an online survey to explore the students' attitude towards the use of the rescaled web-based geographic virtual environment and received additional feedback from the students. This survey included six Likert-scale questions related to the motivations of our research study, in other words, on the (perceived) potentials of the environment to:

- Help the students to develop a better understanding of the project site
- Be used by the students of the design studio as an information resource in the future
- Be used by other students as an information resource in the future
- Support learning from other students
- Motivate collaborative work
- Improve the quality of the group designs

According to the results of the survey study, 87% of the students strongly, mostly or somewhat agreed that using the GEO-VEM had helped them to ***develop a better understanding of Brussels*** (Figure 10, on the left). This observation is important because it illustrates a possible added value of the GEO-VEM in the eyes of the students. When the students were asked if they plan to ***use the virtual environment (GEO-VEM) as an information resource in the future***, 84% of the students responded positively (Figure 10, on the right). Similarly, 88% percent of the students also strongly, mostly or somewhat agreed that ***other students (different than their classmates) can use the environment as an information resource in the future***. These findings relate to the students' positive perception of one of the most important potentials of the Design Studio 2.0, the possibility of transferring design knowledge constructed through the design studio to concurrent and future design studios, design practices and design researchers.

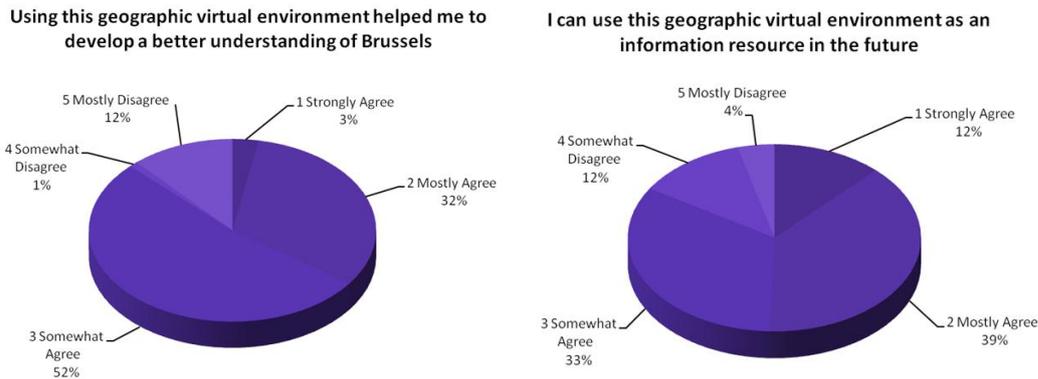

*FIG. 10: Students' responses to two of the Likert-scale questions presented in our research.*

In addition, concerning potentials of the environment to **support learning from other students**, the participants' attitude was also promising: 84% strongly, mostly or somewhat agreed that they learned from their peers. This result is in line with the positive perception of the 76% of the students towards the potentials of the virtual environment to **motivate collaborative work**. The last survey question was on the perceived potentials of virtual environment to **improve the quality of the group designs**. As we have detected a possible correlation between collaborative edits and student performance in the previous section, the responses of the students could possibly point towards a positive contribution to the design that they have created as a group. In contrast, we observed that the students' positive responses to this question were the lowest among all of the questions: only 56% strongly, mostly or somewhat agreed that using the GEO-VEM improved the quality of their group designs.

At the end of our online survey, we also asked an open-ended question related to the future development of our online environment to receive the comments and suggestions from the students. We have received many encouraging responses (Table 2) as well as critical remarks on the general and specific aspects of the virtual environment.

*Table 2: Samples from the positive student responses to the open-ended question in our survey.*

| Do you have any thoughts for the future development of the virtual environment? |
|---|
| *Student A: "The website itself was very easy to use, I haven't made a website before but felt it very easy to understand how to do it and have now basic knowledge I can use later. I think the interface should be as easy/simple as now to make everyone able to understand and use it."* |
| *Student B: "I'm not really good with computers and the idea of "virtual environment" sounded really abstract for me at the beginning. But I think I really understood and developed an interest to use it as an analysis tool."* |
| *Student C: "It is a nice idea, because our people can see our work. I never thought that I could edit the website; it was an unknown field for me. However it turned out that it isn't that difficult. At the end, in my group, I was the person who edited our page the most"* |
| *Student D: "This tool is quite important for the globalized world we move in nowadays, and it's a powerful way to explain your ideas, readable from wherever and easy to use."* |
| *Student E: "In the future it would be more interesting to learn from other students by checking their own inspirations, creations, ideas, sketches and so on."* |

The critical remarks can be summarized as: the difficulty of learning wiki syntax, system errors due to browser compatibility issues, importance of the regularity of the announcement updates and time consuming nature of scanning and uploading manual drawings to the online environment.

In conclusion, the findings of the online student attitude survey demonstrate a (general) positive attitude towards the use of the GEO-VEM. These results are encouraging for the improvement and future testability of similar Design Studio 2.0 applications, but they are not solely sufficient for evaluation.

In order to test the reliability of the survey, we have presented the same questions to the same students four months later in print format. The comparative analysis indicated a high level of correlation between the two measurements (Pak and Verbeke, 2011). These results can be considered as suggestive evidence for the reliability of the questionnaire as a method. But it is important to note that the questionnaire results should be taken as a reflection of the attitudes of the students rather than the reality itself.

We believe that more reliable answers to these six survey questions can be derived from long-term studies on these kinds of studio experiments. It is expected that the differences in the (designerly) profiles and individual characteristics of the students can potentially have negative and/or positive effects on the results.

Another interesting observation from our design studio experience was the students' reluctance to comment on each other's findings and projects through the web interface (talk pages). Although the groups were able to critically develop their findings and projects on their group theme pages together while reading other groups' works, especially the highly critical discussions between the groups mostly took place in an oral and face to face manner, in the conventional design studio. As a future recommendation related to this topic, combining the social networking environments or chat functionalities with the wiki talk pages may motivate the students to use the online commenting functionalities more frequently.

## 5.3 Post-task Student Satisfaction Survey

After concluding the design studio, we gave eleven basic tasks to 25 students (out of 39 who attended the course) to observe their level of satisfaction. The tasks that they had to accomplish were:

- Logging in to the website
- Searching for a topic (i.e.: Networks, Fragile)
- Creating a topic
- Editing and formatting a page
- Adding a (Google) map to a page
- Adding a placemark on the (Google) map
- Adding an explanation to a (Google) placemark
- Drawing a line on the (Google) map
- Drawing an area on the (Google) map
- Importing an external (Google) map to page
- Placing multiple maps on top of each other

After performing these tasks, the students answered three basic questions on their satisfaction levels, regarding their (perceived) *ease of completion*, *amount of time it required to accomplish the tasks* and the *provided support information* (help documentation). According to the results of this study (Figure 11), 96 % of the students strongly, mostly or somewhat agreed that they were satisfied with the ease of completing the tasks. Moreover, when we asked their level of satisfaction on amount of time it required to accomplish the tasks 80% responded positively.

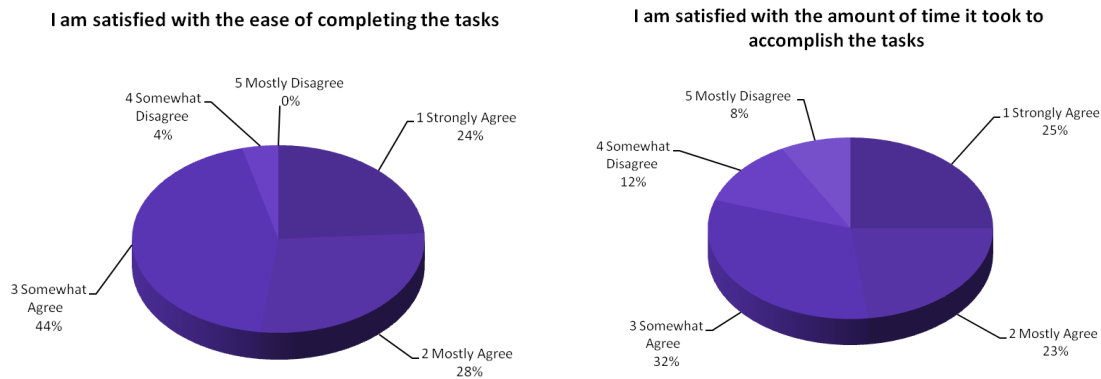

*FIG. 11: Results of the user satisfaction questionnaire.*

The third and final question was about the support information (help documentation) that is provided during the design studio (as reviewed in section 3.2 the help documentation included: tutorials on how to upload and edit text, links and images and create and edit maps and the classical wiki "cheat sheet"). The analysis of the responses to this question showed that 92% of the students strongly, mostly or somewhat agreed that they were satisfied with the help information.

Overall, the analysis results are significant because they indicate a problem/context independent perception of the GEO-VEM by the students; which in this case be regarded as highly positive.

## 6. CONCLUSIONS AND FUTURE DIRECTIONS

In this paper, we have reviewed various educational applications of e-learning strategies in design studios. Considering these applications, we shared our ideas on the potentials of utilizing Web 2.0 software and information aggregation services in a design studio setting. We claimed that with the help of e-learning strategies, it is possible to support, augment and enrich the reflective communication between the design students and studio teachers in a progressive studio setting and named it "Design Studio 2.0" (Section 2.3).

In order to evaluate the Design Studio 2.0 concept in real life, we have rescaled and customized a Web 2.0-based Geographic Virtual Environment Model (GEO-VEM) (Section 3) and incorporated it in a graduate design studio at the Sint-Lucas School of Architecture during the fall semester of 2010 (Section 3.2).

We reported the results of this study in detail including the studio setup, student works and applications of various opportunities provided by the Web 2.0 services in the design studio. We observed that during the whole design studio (2.0), the students actively used our web environment in a reflective manner, and created an impressive online inventory with 66 topics (pages), organized according to 11 themes. These topics included various analysis findings, sketches, photos, maps, studio presentations and texts describing their experiences and thoughts on their future projects (Section 4). This inventory definitely shows the power of social learning and online collaborative analysis.

As reviewed in (Section 5.1), the web analytics revealed that 79% of the pages were edited by more than one student. We can conclude from this observation that the students collaborated through the Web 2.0 environment during the design studio. This observation is also in line with our informal perception that collaborating online in groups motivated them to share information and findings and construct a collective memory of their project sites.

We compared the student grades at the end of week eight with the total number of group edits with the assumption that there could be some kind of a relation between the collaborative use of the environment and group success (grades) in the analysis phase of our design studio. As a result, we found that the groups who received higher grades had made relatively more collaborative edits. It is not possible to derive a direct causality out of this finding but the data points indicate a possible correlation between collaborative edits and student performance.

The results of the student attitude survey illustrated a general positive attitude towards the use of the virtual environment (Section 5.2). The students were convinced that the GEO-VEM helped them to develop a better understanding of the project site. They expressed that they wanted to use it as an information resource in the future, and it supported learning from other students and motivated them to work online collaboratively. These results are encouraging for the further development and future testability of similar Design Studio 2.0 applications, but they are not solely sufficient for the evaluation of the environment. More reliable answers can be derived from long-term studies on Design Studio 2.0 experiments. It is expected that the differences in the profiles and individual characteristics of the students can potentially have negative and/or positive effects on the results.

The post-task student satisfaction questionnaire (Section 5.3) also showed high levels of satisfaction among the studio participants. These analysis results were significant because they have indicated a positive perception of the GEO-VEM by the students independent from the problem and the context.

For the future development of the Web 2.0-based GEO-VEM, we asked an open-ended question to the students to receive the comments and suggestions. We received many encouraging responses as well as critical remarks on the general and specific aspects of the virtual environment. The critical remarks can be summarized as: the difficulty of learning wiki syntax, system errors due to browser compatibility issues, importance of the regularity announcement updates and time consuming nature of scanning and uploading manual drawings to the online environment.

We also individually experienced the usability problems related to the lack of an efficient What You See Is What You Get editing interface, the (relative) complexity of the wiki and keyhole mark-up language for the students. Because of these problems, the semantic extension was not effectively utilized by the students, which points out to a missed opportunity. We believe that it is possible to overcome the first two challenges by using a more sophisticated editing tool and providing students with basic information on mark-up languages. Moreover, in the near future, with the development of affordable and portable multi-touch devices and tablets, students can create their drawings in digital format which can make them easier to share with other students.

In addition, the presence of the conventional and online learning at the same time (blended learning) helped us to stretch the limits of the web environment. Learning how to use the environment and processing information to be represented online were challenging tasks for some of the students and teaching staff; especially at the beginning of the design studio. In this context, weekly (physical) workshops were highly useful for raising awareness on the conceptual framework of the implementation of the virtual environment as well as solving technical issues.

Overall, together with the formal surveys, our informal observations suggest that the students learned to communicate and reflect on their designs using various means, including alternative analysis topics, images and models, all of which stimulate them to think more about the conceptual foundations of their projects. The design studio coordinators noted that this process has induced more competition between the students.

In addition, the proposed GEO-VEM provided opportunities for the transfer of the rich knowledge produced within the framework of a design studio to future studios, thus establishing a basis for the sustainable development of education and design ideas. The design studio coordinators were also assured that the body of knowledge represented in the GEO-VEM can potentially inspire their future students, and therefore we decided to use this environment as a major resource for next year's design studios. During the time of preparation of this paper, it was being successfully used in a new urban design studio setting.

In conclusion, we learned from this study that creating virtually transparent and open studios can enhance the communication in architectural design education. The virtual environment that we tested in the proposed design studio context performed as a reliable information platform for collecting and disseminating students' design information and concepts and motivated them to collaborate. We were also able to use the environment for following the progress of student works online on a regular basis, especially during the reflection process which took place in the design studio.

## 7. ACKNOWLEDGEMENTS

This research was supported by a three year type (B) "Prospective Research for Brussels" postdoctoral research grant from the Brussels Capital Regional Government, Institute for the Encouragement of Scientific Research and Innovation (INNOVIRIS) given to dr. Burak Pak, promoted by Prof. dr. Johan Verbeke.

The OIKODOMOS project is funded by the European Commission Lifelong Learning Program - Erasmus Virtual Campus, Grant Reference: 134370-LLP-1-2007-1-ES-ERASMUS-EVC

## 8. REFERENCES

Brown, J. Adler, R. P. (2008). Minds on Fire: Open Education, the Long Tail, and Learning 2.0, *Educause Review* (January/February 2008), 16-32.

Burrow, A., Burry, J,. (2006). Working with Wikis, by Design. In, *Architectural Design, Special Edition: Collective Intelligence in Design* (Castle, H., Perry C. and Hight C., editors): Academy Editions, John Wiley and Sons.

Chase S., Schultz R, Brouchoud, J (2008). Gather around the Wiki-Tree: Virtual Worlds as an Open Platform for Architectural Collaboration, *Architecture in Computro: Integrating methods and techniques, Proceedings of the 26th Conference on Education in Computer Aided Architectural Design in Europe* (Muylle, M., editor), Antwerp: Artesis University College of Antwerpen, 809-815

EAAE member profiles, accessed 20 March 2012 <http://www.eaae.be/members_new.php>

Glott, R., Schmidt, P., & Ghosh, R. A. (2010). Wikipedia survey—Overview of results. Retrieved from UNI-MERIT, accessed 20 March 2012 <http://www.wikipediasurvey.org/docs/Wikipedia_Overview_15March2010-FINAL.pdf >

Heinze, A., Procter, C. (2004). Reflections on the Use of Blended Learning, *Education in a Changing Environment International Conference Proceedings* (Doherty, E. editor),University of Salford, Salford:Informing Science,

Hudson-Smith, A., Batty, M., Crooks, A. T., and Milton R (2009). Mapping Tools for the Masses: Web 2.0 and Crowdsourcing, *Social Science Computer Review*, vol. 27 no.4, pp. 524-538.

Jacques A. (1982). The Programmes of the Architectural Section of the Ecole Des Beaux Arts, 1819-1914, in The Beaux Arts and 19th Century French Architecture (Middleton, R., editor). London: Thames & Hudson.

Jonassen, D. H. (1999). Constructing learning environments on the web: Engaging students in meaningful learning. EdTech *99: Educational Technology Conference and Exhibition* 1999: Thinking Schools, Learning Nation.

Lackey, J. A (1999). A history of the Studio based Learning Model, Mississippi State University Educational Design Institute accessed 20 March 2012 <http://www.edi.msstate.edu/work/pdf/history_studio_based_learning.pdf>

Lindquist, M. (2006). Web Based Collaboration (for Free) Using Wikis in Design Studios *ACADIA 2006: Synthetic Landscapes Digital Exchange Conference,* University of Kentucky October 12-15, 2006 Louisville, Kentucky.

Madrazo, L., Riddy P., (2011). OIKODOMOS Virtual Campus: Constructing learning processes in collaboration, *Education and Research in Computer Aided Architectural Design in Europe (eCAADe) Conference (Zupancic, T., Juvancic, M,, Verovsek, S., Jutraz, A., editors),* University of Ljubljana (UL), Slovenia.

Madrazo, L., Riddy P., Sicilio A. (2010). OIKODOMOS Technological Platform, *EDULEARN International Conference on Education and New Learning Technologies*, Barcelona, 5-7 July 2010.

OIKODOMOS project, accessed 20 March 2012 <http://www.oikodomos.org>

Pak, B. (2009). A Virtual Environment for Analysis and Evaluation of Alternative Urban Development Projects for the Brussels Capital Region, *Institute for the encouragement of Scientific Research and Innovation of Brussels, Prospective Research for Brussels Project* 2009.

Pak, B., Kuhk, A. (2009). Re-inventing Brussels: how knowledge on alternative urban development projects can alter urban policies, *The 4th International Conference of the International Forum on Urbanism (IFoU), The New Urban Question*, 2009. Amsterdam/Delft, 1421.

Pak, B., Verbeke, J. (2011). Usability as a Key Quality Characteristic for Developing Context-friendly CAAD Tools and Environments*, Education and Research in Computer Aided Architectural Design in Europe*

*(eCAADe) Conference (Zupancic, T., Juvancic, M,, Verovsek, S., Jutraz, A., editors),* University of Ljubljana (UL)*,* Slovenia, 275-276.

Puntambekar, S. and Young, M.F. (2003). Moving towards a theory of CSCL, Designing for Change in Networked Learning Environments (Wasson, B., Ludvigsen, S. and Hoppe, U., editors.), London:Kluwer Academic Publishers, 503-512.

Schön, D. (1983). The reflective practitioner: How professionals think in action. New York: Basic Books.

Schön, D. (1986). The design studio: An exploration of its traditions and potential. London: Royal Institute of British Architects.

Schön, D. (1987). Educating the reflective practitioner: Toward a new design for teaching and learning in the professions. San Francisco: Jossey-Bass.

Shirky, C. (2010). Cognitive Surplus: Creativity and Generosity in a Connected Age, Penguin Press HC.

Turk, Z. (1991). Integration of Existing Programs Using Frames, *CIB Seminar Computer Integrated Future*, 16-17 September, Eindhoven, Netherlands, accessed 20 March 2012 <http://www.zturk.com/data/works/att/f88c.fullText.pdf >

Verbeke, J., Ooms T., Madrazo L. and Riddy Paul (eds.) (2012) OIKODOMOS, Brussels: Hogeschool voor Wetenschap & Kunst.

Vitruvius, P., (transl. Morris Hicky Morgan, 1914) The Ten Books on Architecture. Neeland Media LLC .

Young, R. M. (1994) Mental Space, Process Press Ltd, London, accessed 20 March 2012<http://human-nature.com/rmyoung/papers/paper55.html>